\documentclass[aps,prl,twocolumn,superscriptaddress,groupedaddress]{revtex4}  
\usepackage{graphicx,comment,float,braket,textcomp,amsmath,lipsum,color,soul}  
\usepackage{dcolumn}   
\usepackage{bm}        
\usepackage{amssymb}   

\begin{document}

\title{Partially-Bright Triplet Excitons in Perovskite Nanocrystals}

\author{A. Liu} \thanks{These authors contributed equally.} \affiliation{Department of Physics, University of Michigan, Ann Arbor, Michigan 48109, USA}

\author{D. B. Almeida} \thanks{These authors contributed equally.} \affiliation{Department of Physics, University of Michigan, Ann Arbor, Michigan 48109, USA}

\author{L. G. Bonato} \affiliation{Instituto de Quimica, Universidade Estadual de Campinas, 13083-970 Campinas, Sao Paulo, Brazil}

\author{G. Nagamine} \affiliation{Instituto de Fisica ``Gleb Wataghin", Universidade Estadual de Campinas, 13083-970 Campinas, Sao Paulo, Brazil}

\author{L. F. Zagonel} \affiliation{Instituto de Fisica ``Gleb Wataghin", Universidade Estadual de Campinas, 13083-970 Campinas, Sao Paulo, Brazil}

\author{A. F. Nogueira} \affiliation{Instituto de Quimica, Universidade Estadual de Campinas, 13083-970 Campinas, Sao Paulo, Brazil}

\author{L. A. Padilha} \email{padilha@ifi.unicamp.br} \affiliation{Instituto de Fisica ``Gleb Wataghin", Universidade Estadual de Campinas, 13083-970 Campinas, Sao Paulo, Brazil} 

\author{S. T. Cundiff} \email{cundiff@umich.edu} \affiliation{Department of Physics, University of Michigan, Ann Arbor, Michigan 48109, USA}

\vskip 0.25cm

\date{\today}

\begin{abstract}
Advances in opto-electronics require the development of materials with novel and engineered characteristics. A class of materials that has garnered tremendous interest is metal-halide perovskites, stimulated by meteoric increases in photovoltaic efficiencies of perovskite solar cells. In addition, recent advances have applied perovskite nanocrystals (NCs) in light-emitting devices. It was discovered recently that, for cesium lead-halide perovskite NCs, their unusually efficient light-emission may be due to a unique excitonic fine-structure composed of three bright triplet states that minimally interact with a proximal dark singlet state. To study this fine-structure without isolating single NCs, we use multi-dimensional coherent spectroscopy at cryogenic temperatures to reveal coherences involving triplet states of a CsPbI$_3$ NC ensemble. Picosecond timescale dephasing times are measured for both triplet and inter-triplet coherences, from which we infer a unique exciton fine-structure level-ordering comprised of a dark state energetically positioned within the bright triplet manifold. 
\end{abstract}

\maketitle

\section{Introduction}

Cesium lead-halide perovskites were first synthesized over a century ago with a general chemical formula CsPbX$_3$ (where X = Cl, Br, or I). Recently, synthesis of CsPbX$_3$ nanocrystals (NCs) was achieved \cite{Protesescu2015,Ha2017}, which combines the advantages of perovskites (e.g., efficient luminescence, long carrier diffusion length) with that of colloidal NC materials (e.g., surface engineering, size-tunable emission). Perovskite NCs exhibit luminescence with quantum yields reaching nearly unity \cite{Liu2017}, in contrast to the optimized 80\% quantum yield achieved by chalcogenide NCs coated with a gradient shell \cite{Jaehoon2014}. Although all other colloidal materials suffer inhibited emission from lower energy dark states \cite{Nirmal1995}, the unusual brightness of perovskite NCs is now believed to originate from an optically active, nondegenerate triplet state that emits efficiently despite the presence of a dark singlet state \cite{Becker2018,Tamarat2019}.

The unique exciton fine-structure of perovskite NCs has significantly extended the potential applications of colloidal NCs. In particular, the three non-degenerate bright triplet states and their orthogonally-oriented dipole moments have generated much excitement for potential applications in quantum information processing \cite{Rainò2018,Utzat2019}. However, engineering exciton superposition states as information carriers will require an intimate knowledge of their coherent dynamics, which are still not well-understood. The exciton fine structure of perovskite NCs has thus far only been studied via single-NC photoluminescence \cite{Yin2017,Becker2018,Chen2018} and transient absorption \cite{Makarov2016,Castañeda2016} techniques, which have provided information only about their incoherent population dynamics. Furthermore, inhomogeneous spectral broadening due to NC size dispersion limits the utility of linear spectroscopic techniques in studying NC ensembles. More sophisticated methods are thus required \cite{Moody2015,Hao2016,Liu2019,CdSeAcousticPhononPaper} to extract the desired ensemble-averaged coherent properties of perovskite NCs. 

\begin{figure*}[t]
    \centering
    \includegraphics[width=0.95\textwidth]{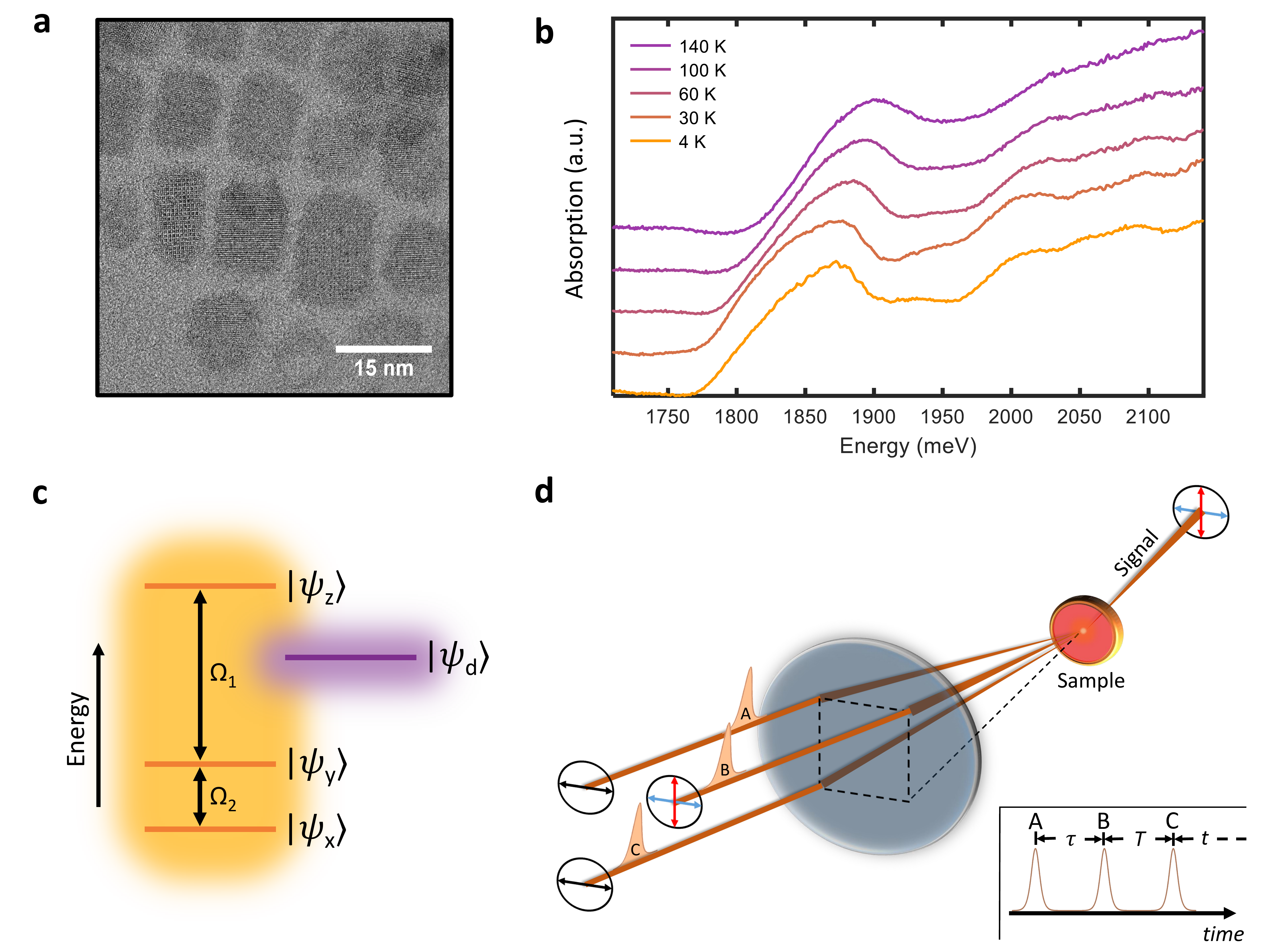}
    \caption{{\bf CsPbI$_3$ perovskite nanocrystals studied via MDCS.} {\bf a}, Transmission electron micrograph of representative CsPbI$_3$ NCs. {\bf b}, Perovskite NC absorption spectra as a function of temperature \cite{ThreeOscillatorPaper}. {\bf c}, Energy level diagram of the non-degenerate bright triplet states $\{\psi_x,\psi_y,\psi_z\}$ that form the band-edge. The dark singlet state $\Ket{\psi_d}$ is shown to lie between states $\Ket{\psi_y}$ and $\Ket{\psi_z}$, which is argued in the main text. {\bf d}, Schematic of the MDCS experiment. Three pulses A, B, and C arranged in the box geometry are focused onto the sample with varying time delays as shown in the inset. Double-sided arrows in circles denote the polarization of each pulse. Pulses A and C are horizontally polarized, which is indicated by the horizontal arrows. Pulse B is either horizontally or vertically polarized, which corresponds to an emitted signal of either horizontal or vertical polarization respectively as indicated by arrows of matching color of the emitted signal.}
    \label{Fig1}
\end{figure*}

Here, we extract crucial figures of merit for quantum information processing, the ensemble-averaged triplet coherence times, and reveal coherence times of both optical triplet coherences and terahertz inter-triplet coherences for an ensemble of CsPbI$_3$ NCs (see Fig. \ref{Fig1}a-b) at cryogenic temperatures. We also present evidence for a mixed bright-dark level-ordering (see Fig. \ref{Fig1}c) that renders the triplet state excitons only partially bright. These measurements are enabled by using multi-dimensional coherent spectroscopy (MDCS) \cite{Cundiff2013} to circumvent the inhomogeneous broadening that obscures spectral signatures of the exciton fine-structure and resolve coherent coupling involving different triplet states. The extracted coherence times for CsPbI$_3$ NCs are an order-of-magnitude longer compared to candidate materials for valley-tronics \cite{Moody2015,Hao2016,Schaibley2016}, which positions perovskite NCs as a potential material platform for quantum information applications via bottom-up assembly.

\begin{figure*}[t]
    \centering
    \includegraphics[width=0.95\textwidth]{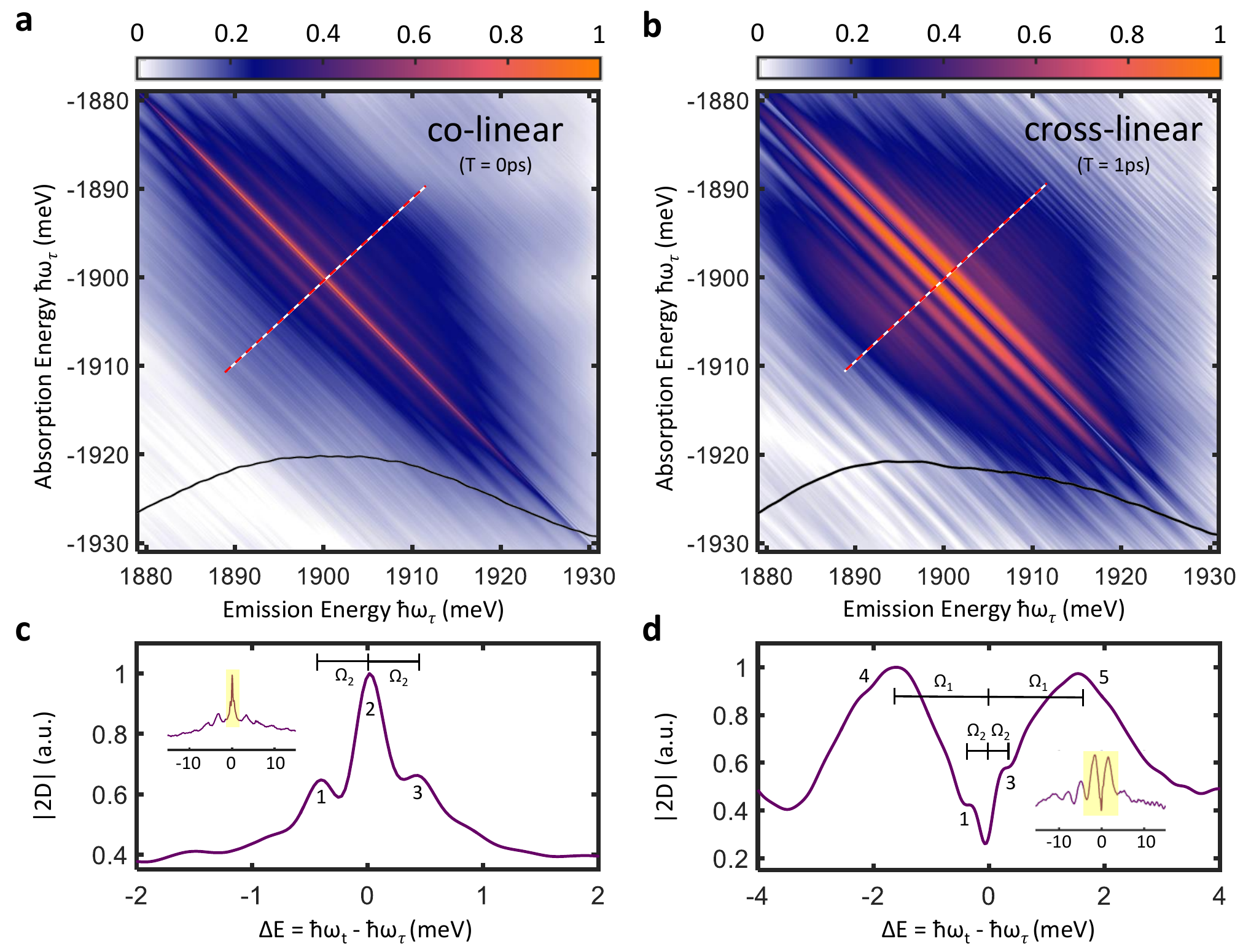}
    \caption{{\bf Triplet coherences in one-quantum spectra.} Magnitude one-quantum spectrum at 4.6 K with {\bf a}, co-linear and {\bf b}, cross-linear excitation. The white/red dashed lines and solid black lines indicate the cross-slice locations and laser pulse spectra respectively. Cross-slice centered at $|\hbar\omega_\tau| = |\hbar\omega_t| = 1900$ meV of the {\bf c}, co-linear and {\bf d}, cross-linear excitation one-quantum spectrum. Numbers in {\bf c} and {\bf d} indicate peaks arising from electronic interband coherences and populations.}
    \label{Fig2}
\end{figure*}

To perform MDCS we use a Multi-Dimensional Optical Nonlinear Spectrometer \cite{Bristow2009}, which focuses three laser pulses onto the perovskite NC ensemble sample as a function of three time delays $\tau$, $T$, and $t$ (schematically shown in Fig. \ref{Fig1}d and inset). By Fourier transforming the emitted four-wave mixing (FWM) signal as a function of two or all three time delays, the coherences and populations induced by each pulse are correlated in a multi-dimensional spectrum. In this study, we Fourier transform along the variables $\tau$ and $t$ to obtain one-quantum spectra (which correlate the absorption energy $\hbar\omega_\tau$ with the emission energy $\hbar\omega_t$) and along the variables $T$ and $t$ to obtain zero-quantum spectra (which correlate the intraband mixing energy \cite{Yang2008} $\hbar\omega_T$ and the emission energy $\hbar\omega_t$). Furthermore, the polarization of the second pulse (labeled B in Fig. \ref{Fig1}d) is chosen to align either parallel or orthogonal to the co-linear polarizations of the other two pulses to probe different quantum pathways. We denote the two polarization schemes as co-linear excitation and cross-linear excitation respectively.

\section{One-Quantum Spectra Probe Optical Triplet Coherences}

One-quantum spectra were acquired at a temperature of 4.6 K with co-linear and cross-linear excitation (shown in Figs. \ref{Fig2}a and \ref{Fig2}b). Both spectra show numerous peaks that are elongated in the diagonal direction ($|\hbar\omega_\tau| = |\hbar\omega_t|$), reflecting inhomogeneous broadening \cite{Siemens2010}. By taking cross-diagonal slices (indicated by the red/white dashed lines in Figs. \ref{Fig2}a and \ref{Fig2}b), the ensemble-averaged homogeneous response of NCs with a certain resonance energy is obtained \cite{Siemens2010}. We plot cross-diagonal slices of the one-quantum spectra in Figs. \ref{Fig2}c and \ref{Fig2}d. In the full slices (inset), asymmetric peaks are observed for $|\Delta E| \gtrsim 4$ meV which we attribute to electronic-vibrational coupling. The main plots of each slice section (highlighted by the yellow boxes inset) show symmetric peaks that, due to their polarization dependence, we attribute to absorption and emission involving different triplet state coherences.

In the third-order perturbative limit, the origin of observed peaks is interpreted as changes in the system density matrix induced by each pulse that form accessible quantum pathways \cite{Mukamel1999}. 
Peaks 1 and 3 in Fig. \ref{Fig2}c are generated by absorption and emission of coherences involving $\Ket{g}$ and both triplet states $\Ket{\psi_x}$ and $\Ket{\psi_y}$. We note that our measurements do not inform the ordering of states $\Ket{\psi_{x/y}}$, so we assume the ordering shown in Fig. \ref{Fig1}c for labeling the dephasing rates discussed below. The central peak 2 is likewise generated by quantum pathways involving absorption and emission by coherences of identical resonance energy $\Ket{g}\Bra{\psi_i}$ and $\Ket{\psi_i}\Bra{g}$ respectively. In Fig. \ref{Fig2}d peaks 1 and 3 are visible as shoulders on two new peaks 4 and 5, which are generated by absorption and emission of coherences involving $\Ket{g}$ and triplet states $\Ket{\psi_y}$ and $\Ket{\psi_z}$. The polarization dependence of all five peaks reflects orthogonally-oriented linear dipole moments of the three triplet states. Here, the peak strengths are determined by the projection of each dipole moment onto the observation plane of each NC. However, the absence of a clear sideband at $\Delta E = \pm (\Omega_1 + \Omega_2)$, corresponding to coupling between the transitions involving $\Ket{g}$ and triplet states $\Ket{\psi_x}$ and $\Ket{\psi_z}$, suggests a much stronger dipole moment for $\Ket{\psi_y}$ compared to those of $\Ket{\psi_x}$ and $\Ket{\psi_z}$. 

Fitting the cross-diagonal lineshapes also extracts the homogeneous linewidths $\gamma_i$ \cite{Siemens2010} of triplet state transitions between $\Ket{g}$ and $\Ket{\psi_i}$. In this context one-quantum spectra are particularly useful when compared to integrated FWM techniques \cite{Becker2018-2}, since the cross-diagonal slice position $|\hbar\omega_{t}| = |\hbar\omega_\tau|$ reflects an effective NC size (see SI). Though the peaks 1 and 4 and peaks 3 and 5 would ideally be mirror-images, vibrational coupling distorts the lineshapes of peaks 1 and 4. We thus fit only the $\Delta E \geq 0$ side (details in SI). Fit of the co-linear slice lineshape (Fig. \ref{Fig2}c) gives a sideband (peaks 1 and 3) dephasing rate $\hbar\frac{\gamma_x + \gamma_y}{2} = 0.12$ meV (5.49 ps) and a zero-phonon line (peak 2) dephasing rate $0.124$ meV (5.32 ps). Fit of peaks 4 and 5 in the cross-linear slice lineshape (Fig. \ref{Fig2}d) gives a dephasing rate $\hbar\frac{\gamma_y + \gamma_z}{2} = 0.496$ meV (1.33 ps). The fitted triplet state energy splittings are likewise $\Omega_1 = 1.82$ meV and $\Omega_2 = 0.24$ meV. If the dipole moment of state $\Ket{\psi_y}$ is indeed much larger than those of $\Ket{\psi_x}$ and $\Ket{\psi_z}$, the zero-phonon line dephasing rate will approximately equal $\hbar\gamma_y$ which in turn determines the individual triplet state dephasing rates $\gamma_x = 0.116$ meV ($T_2^x = 5.68$ ps), $\gamma_y = 0.124$ meV ($T_2^y = 5.32$ ps), and $\gamma_z = 0.868$ meV ($T_2^z = 0.76$ ps). 


\begin{figure}
    \centering
    \includegraphics[width=0.5\textwidth]{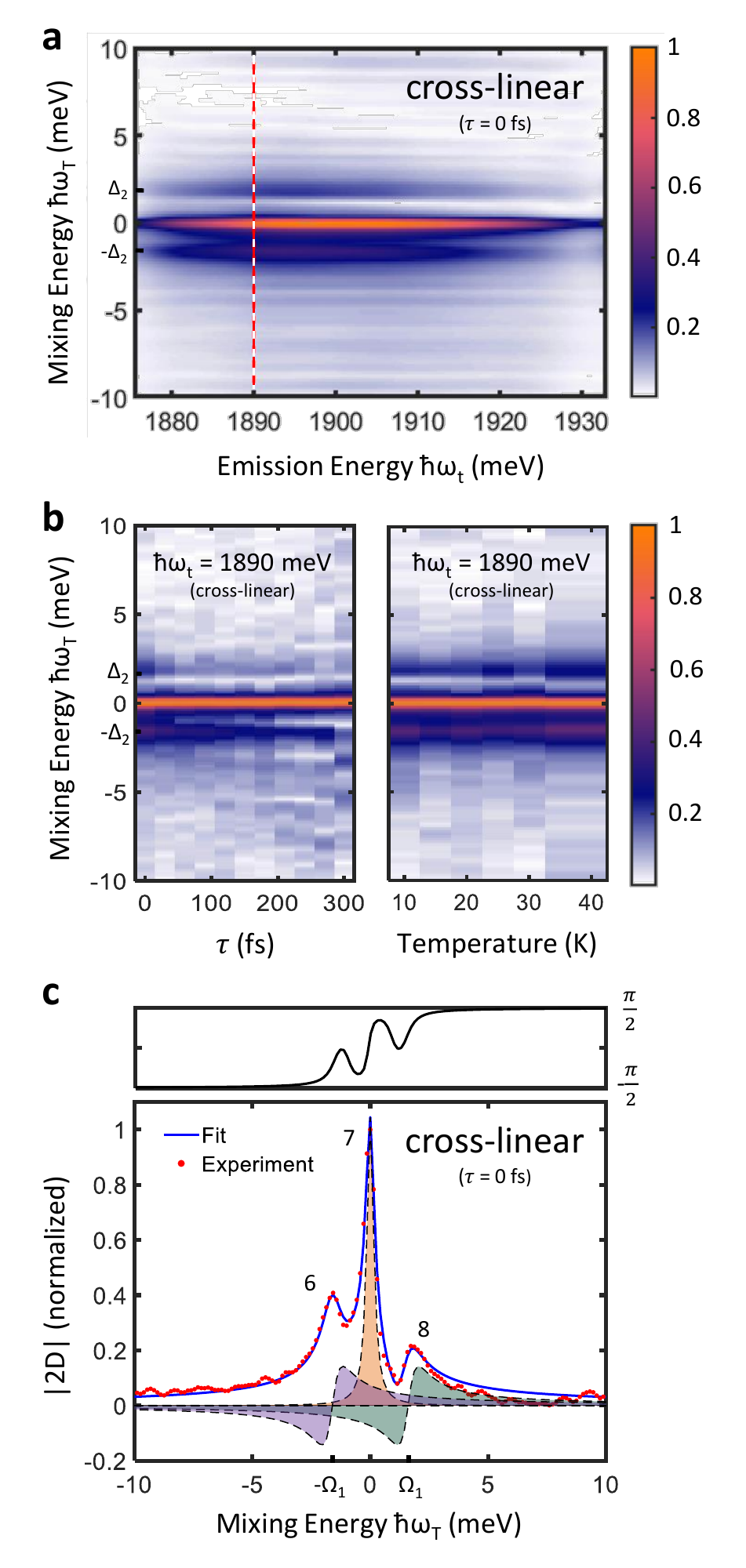}
    \caption{{\bf Inter-triplet coherences in zero-quantum spectra.} {\bf a}, Magnitude zero-quantum spectrum taken at $\tau = 0$ fs and 20 K by passing the FWM signal through a vertical polarizer. Two sidebands due to inter-triplet coherences are observed. {\bf b}, Evolution of normalized slices taken along the dashed red line in {\bf a} at $\hbar\omega_t = 1890$ meV as a function of delay $\tau$ (at 20 K) and temperatures [10, 15, 20, 25, 30, 40] K  (at $\tau = 0$ fs). {\bf c}, Fit of cross-slice taken at $\tau = 0$ fs, in which the complex Lorentzians of peaks 6 and 8 are shifted by phases $-\frac{\pi}{2}$ relative to peak 7. The shaded curves represent the real quadratures of each Lorentzian used to fit lineshape, and the top plot is the phase of the fitted complex lineshape.}
    \label{Fig4}
\end{figure}

\section{Zero-Quantum Spectra Probe Terahertz Inter-Triplet Coherences}

Many of the quantum pathways that generate the sidebands in Figs. \ref{Fig2}a and \ref{Fig2}b involve inter-triplet coherences, which are quantum coherences between triplet states that are not necessarily dipole coupled \cite{FerrioSteel1998}. Of both fundamental and practical importance is the inter-triplet coherence time, which defines the timescale during which the superposition states involved may be coherently manipulated. Inter-triplet coherences are those density matrix elements of the form $\Ket{\psi_i}\Bra{\psi_j}$ where $i,j = \{x,y,z\}$ and $i \neq j$. To directly measure and characterize these coherences, we take zero-quantum spectra at varying temperature and delay $\tau$. For co-linear excitation no inter-triplet coherences between $\Ket{\psi_x}$ and $\Ket{\psi_y}$ are observed (see SI). It is ambiguous whether their corresponding peaks are weak, or are simply obscured by the linewidth of a central $\omega_T = 0$ peak. For cross-linear excitation, we further isolate the inter-triplet coherence pathways by passing the measured FWM signal through a vertical polarizer. We plot a resultant cross-linear zero-quantum spectrum at 20 K in Fig. \ref{Fig4}a. Sidebands are observed at mixing energies identical to the positions of peaks 4 and 5 in Fig. \ref{Fig2}d, which we attribute to inter-triplet coherences between $\Ket{\psi_y}$ and $\Ket{\psi_z}$. An inter-triplet coherence between $\Ket{\psi_x}$ and $\Ket{\psi_z}$ is observed in neither the co-linear nor cross-linear zero-quantum spectra, which is consistent with a dominant transition dipole of state $\Ket{\psi_y}$ as argued above.

In Fig. \ref{Fig4}b, the evolutions of normalized slices (at $\hbar\omega_T = 1890$ meV) as a function of delay $\tau$ and temperature are shown. The FWM signal dephases rapidly with increasing $\tau$ and results in an equally rapid decrease of sideband visibility, in contrast to the opposite behavior of vibrational intraband coherences (see \cite{Liu2019} and SI). No change in the amplitude ratio between sidebands 6 and 8 is observed as temperature increases, confirming that the state splitting observed indeed belongs to the bright-triplet excited state rather than from thermal filling of higher-lying ground states. We note that the triplet state coherences in one-quantum spectra broaden significantly with increasing temperature, and are not resolved at temperatures above 15 K. In contrast, no significant broadening is observed of the inter-triplet coherence linewidth up to 40 K (see Fig. \ref{Fig4}b), indicating that inter-triplet coherences are robust against thermal dephasing \cite{Moody2015}. 

A slice at $\tau = 0$ fs is plotted in Fig. \ref{Fig4}c, from which we can extract the inter-triplet coherence time. However, the quantum pathways that generate peaks 6 and 8 involve identical dipole moments $\mu_y^2\mu_z^2$, from which we expect equal peak amplitudes contrary to the uneven peaks observed. This difference is due to interference between the three complex Lorentzian lineshapes underlying the overall amplitude lineshape, and the fit to experiment is performed by shifting the phase of each sideband Lorentzian lineshape by identical factors of $-\frac{\pi}{2}$ relative to the central $\omega_T = 0$ peak. From our fit, we extract an energy splitting $\Omega_1 = 1.61$ meV and an inter-triplet coherence time $T_2^{yz} = 1.36$ ps at 20 K.

\section{Discussion}

It is quite unexpected that the optical dephasing rate $\gamma_z$ is so much faster than those of the other two triplet states $\gamma_x$ and $\gamma_y$. Although this disparity suggests a fundamentally different dephasing mechanism for coherences involving state $\Ket{\psi_z}$, photoluminescence of similar orthorhombic perovskite NCs that exhibit triplet state structure reveals similar emission linewidths for all three states of the manifold \cite{Becker2018,Tamarat2019}. We resolve this discrepancy by proposing a unique exciton fine-structure comprised of a dark singlet state $\Ket{\psi_d}$ that lies above the states $\Ket{\psi_x}$ and $\Ket{\psi_y}$, which form the band-edge, while remaining below the third triplet state $\Ket{\psi_z}$ (shown in Fig. \ref{Fig1}c). Rapid relaxation from $\Ket{\psi_z}$ to $\Ket{\psi_d}$ then significantly decreases the population lifetime $T_1^z$, and consequently $T_2^z$ as well \cite{Mukamel1999}. Such a fine-structure has been theoretically predicted \cite{Sercel2019} in certain ranges of NC size due to competition between the Rashba effect and electron-hole exchange interaction. Our hypothesis is further supported by previous photoluminescence studies of CsPbI$_3$ NCs of nearly identical size \cite{Yin2017} which revealed polarized doublets corresponding to $\Ket{\psi_x}$ and $\Ket{\psi_y}$ but did not detect the third triplet state $\Ket{\psi_z}$, whose emission would be quenched by non-radiative relaxation to $\Ket{\psi_d}$ according to our model. In accordance with the predicted size-dependence of the relative dark state energy \cite{Sercel2019}, we also observe an abrupt increase in $T_2^y$ with increasing slice position (see SI) which results from a crossing in energy of $\Ket{\psi_y}$ and $\Ket{\psi_d}$.

To conclude, we have measured and characterized both optical frequency triplet coherences and terahertz frequency inter-triplet coherences. We have also presented evidence of an exciton bandedge whose emission is partially quenched by an intermediate dark state, which contributes important insight into the controversial nature of exciton ground states in different perovskite NC materials \cite{Becker2018,Tamarat2019}. 
As a material still in its infancy, perovskite NCs show promise for applications in opto-electronic devices. Particularly, the minimal thermal broadening of inter-triplet coherences observed here motivates study of applications above cryogenic temperatures. For example, in a triplet-state analogue of valleytronics in two-dimensional materials \cite{Schaibley2016}, superpositions of triplet states could be initialized and read-out with linearly polarized light and coherently manipulated via terahertz radiation \cite{Langer2018} as information carriers.

\vspace{1cm}

\noindent {\bf Acknowledgements} This work was supported by the Department of Energy grant number DE-SC0015782. D.B.A. and G.N. acknowledge support from the Brazilian National Council for Scientific and Technological Development (CNPq). L.A.P. acknowledges support from FAPESP (Project numbers 2013/16911-2 and 2016/50011-7). Research was also supported by LNNano/CNPEM/MCTIC, where the TEM measurements were performed.

\noindent {\bf Author Contributions} A.L. and D.B.A. contributed equally to this work. L.A.P. and S.T.C. conceived the concept. L.G.B. synthesized the perovskite NC sample under supervision from A.F.N. G.N. and L.F.Z. acquired TEM images of the sample and characterized the NC size. A.L. and D.B.A. ran the experiments and acquired the data. A.L. analysed the results and wrote the manuscript. All authors discussed the results and commented on the manuscript at all stages.\\

\noindent {\large \bf Materials}

The NCs studied are cube-shaped, with side lengths of 8.7 $\pm$ 2.6 nm measured from transmission electron microscopy measurements (shown in Fig. \ref{Fig1}a). These sizes are comparable to the CsPbI$_3$ exciton Bohr diameter (12 nm) \cite{Protesescu2015}, and correspond to a room temperature 1S exciton absorption peak centered around 1900 meV (shown in Fig. \ref{Fig1}b). The synthesis method is described in the SI.

\newpage
\begin{figure}
   \includegraphics[width=1\textwidth,page=1]{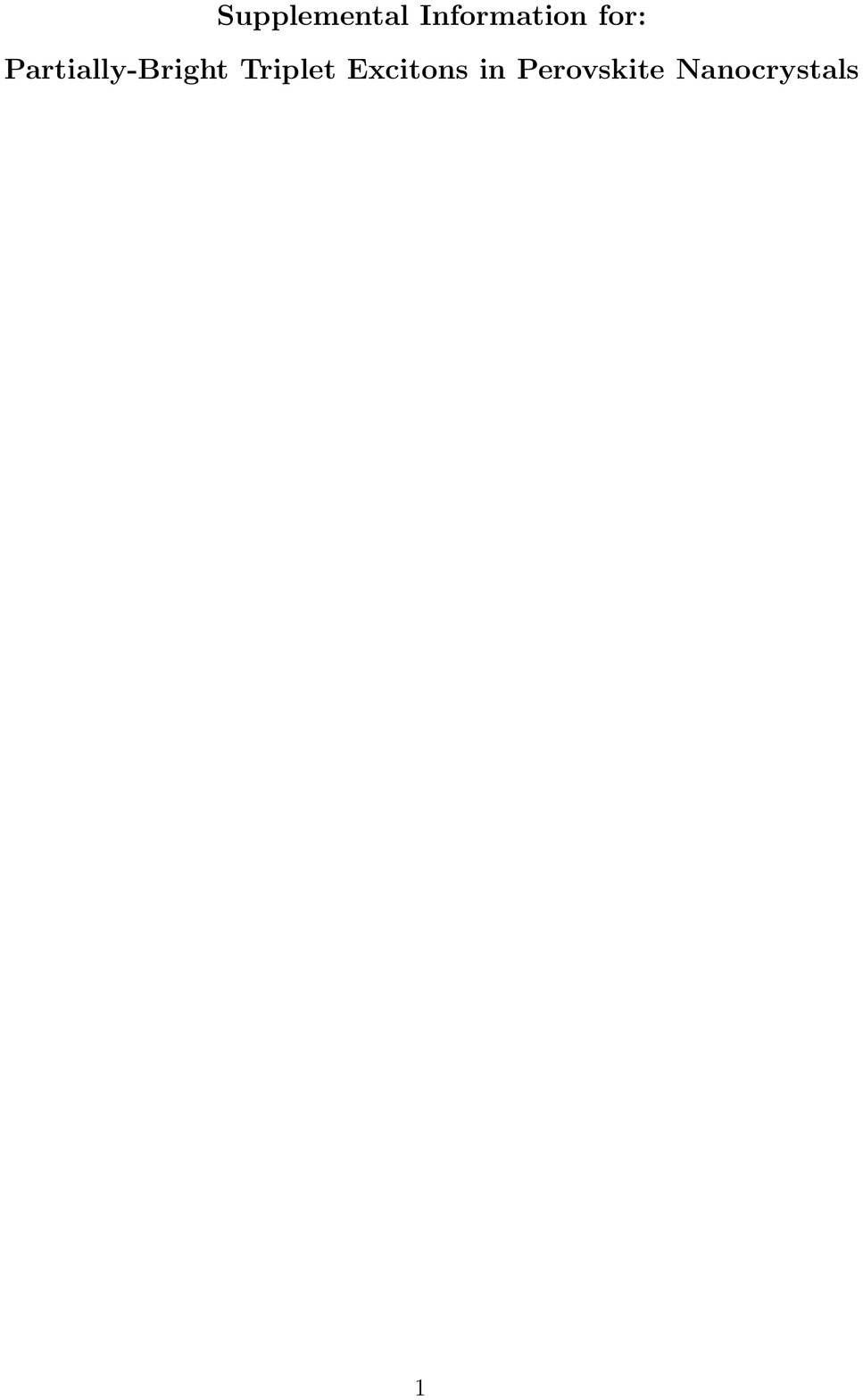}
\end{figure}
\begin{figure}
   \includegraphics[width=1\textwidth,page=2]{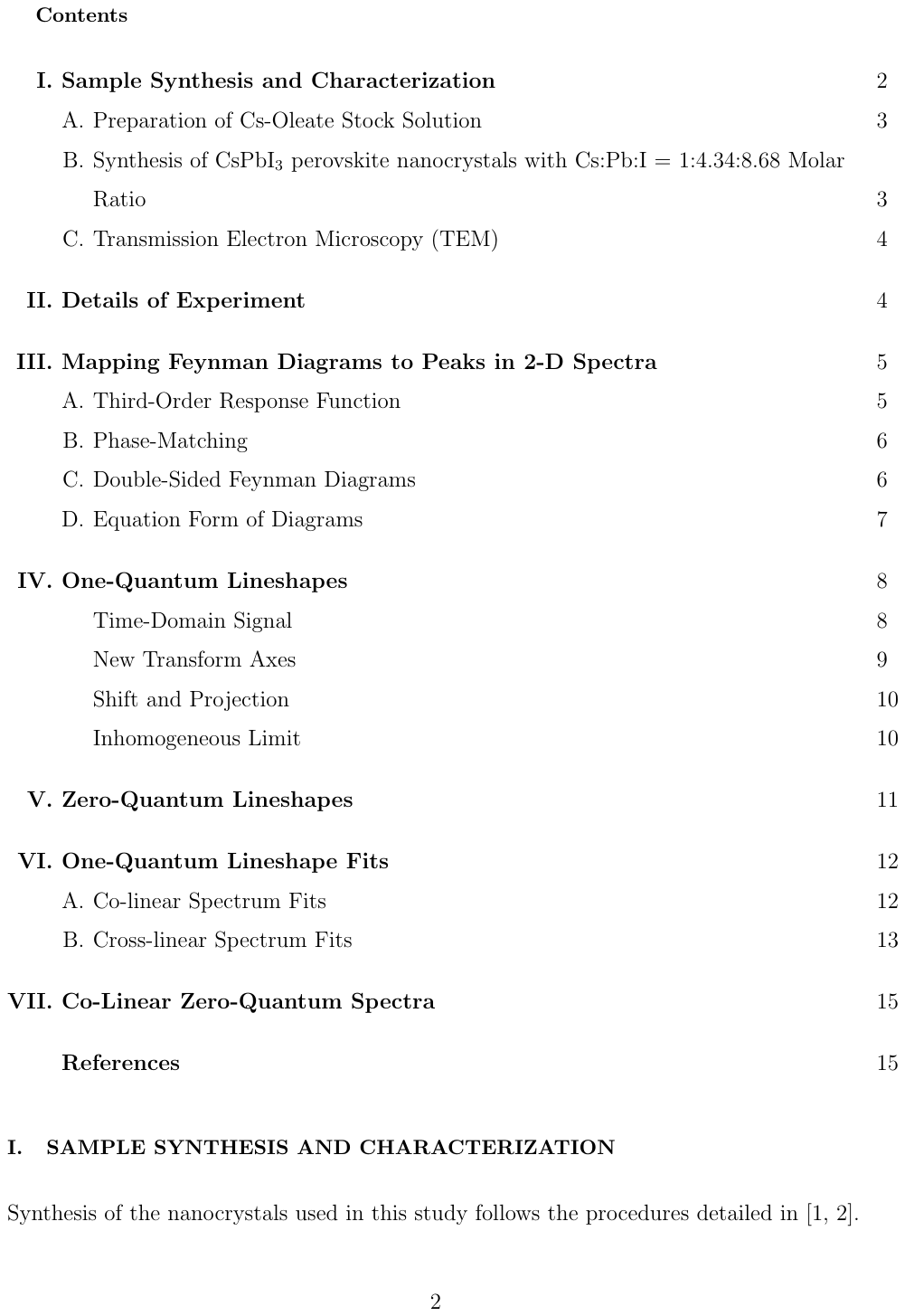}
\end{figure}
\begin{figure}
   \includegraphics[width=1\textwidth,page=3]{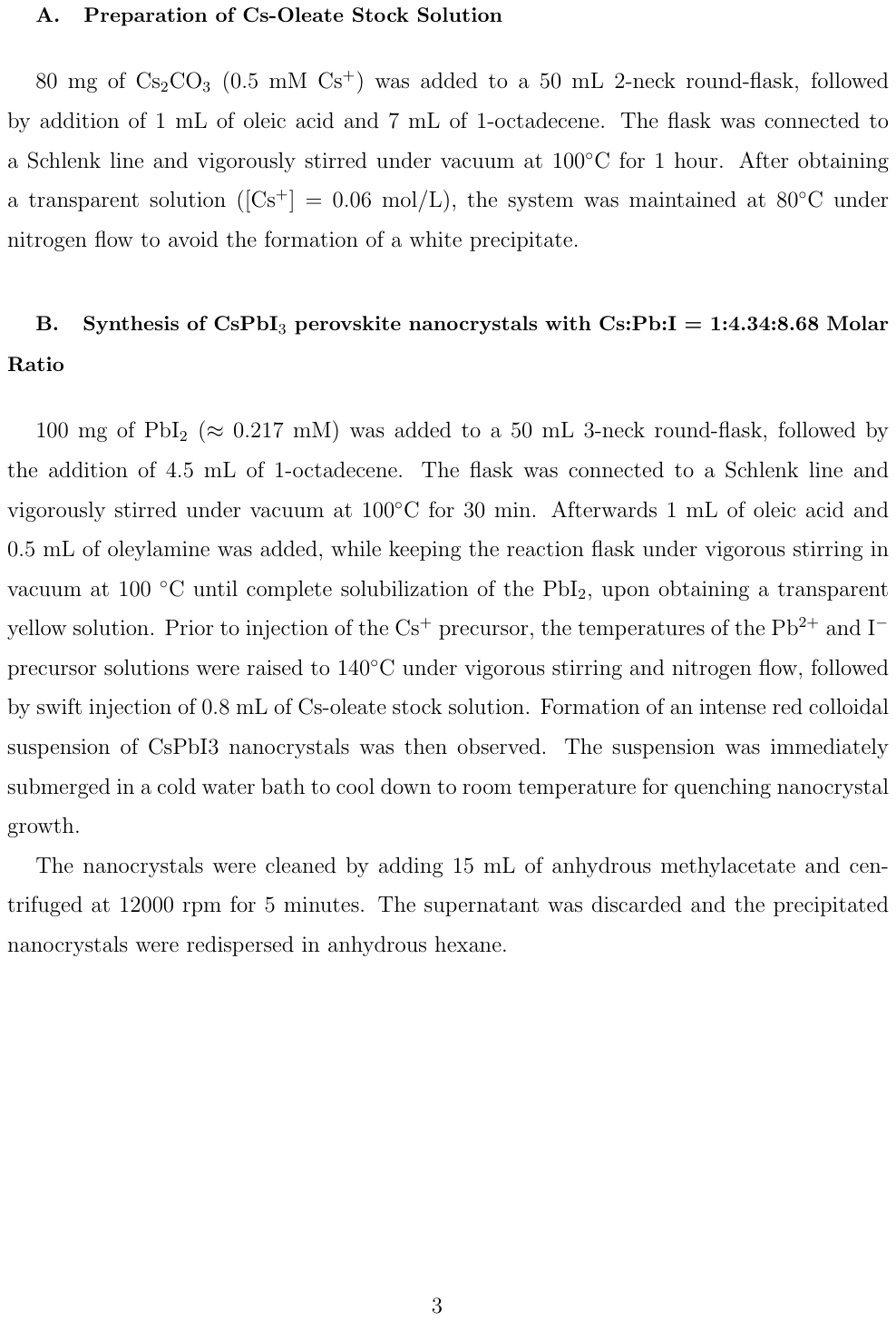}
\end{figure}
\begin{figure}
   \includegraphics[width=1\textwidth,page=4]{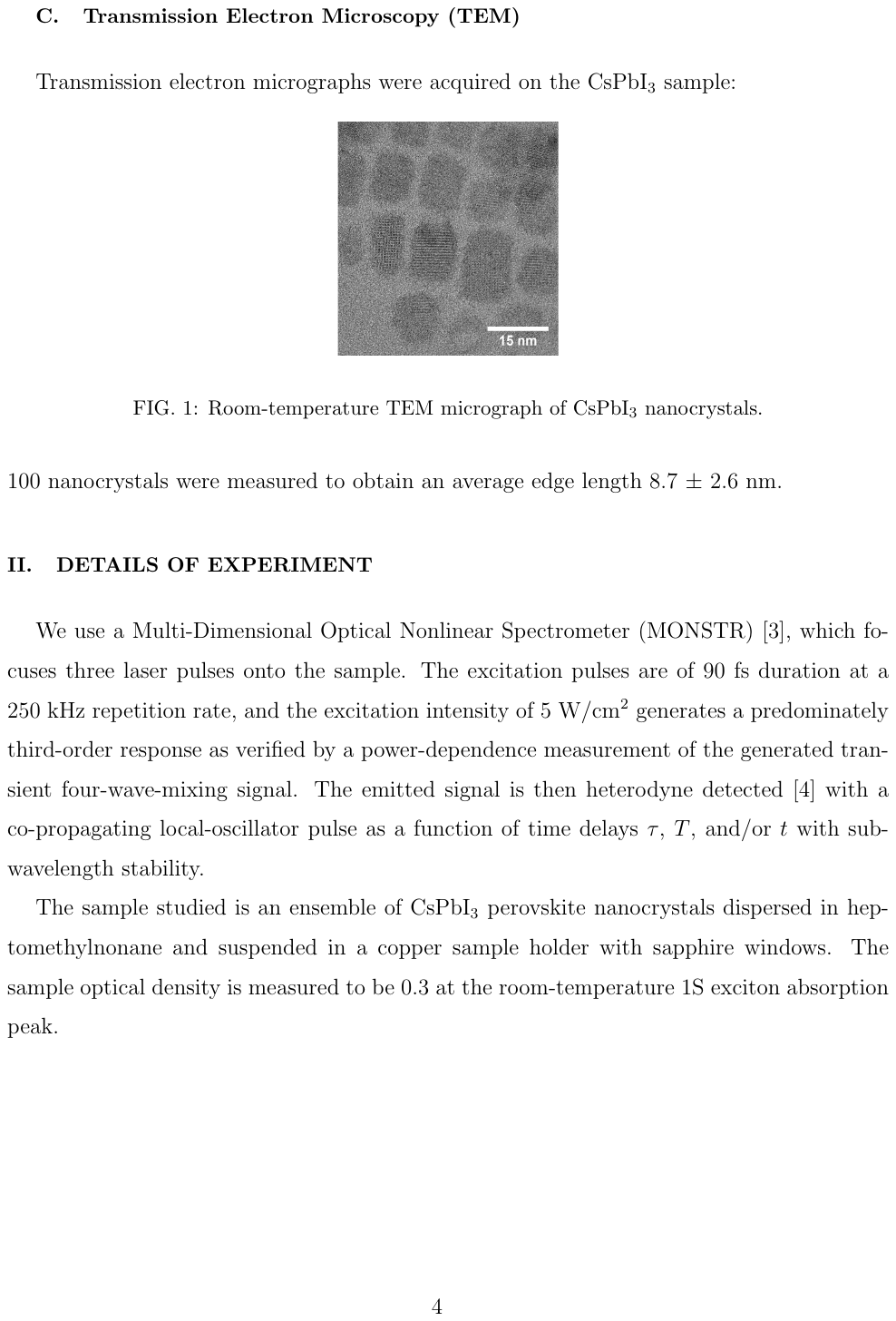}
\end{figure}
\begin{figure}
   \includegraphics[width=1\textwidth,page=5]{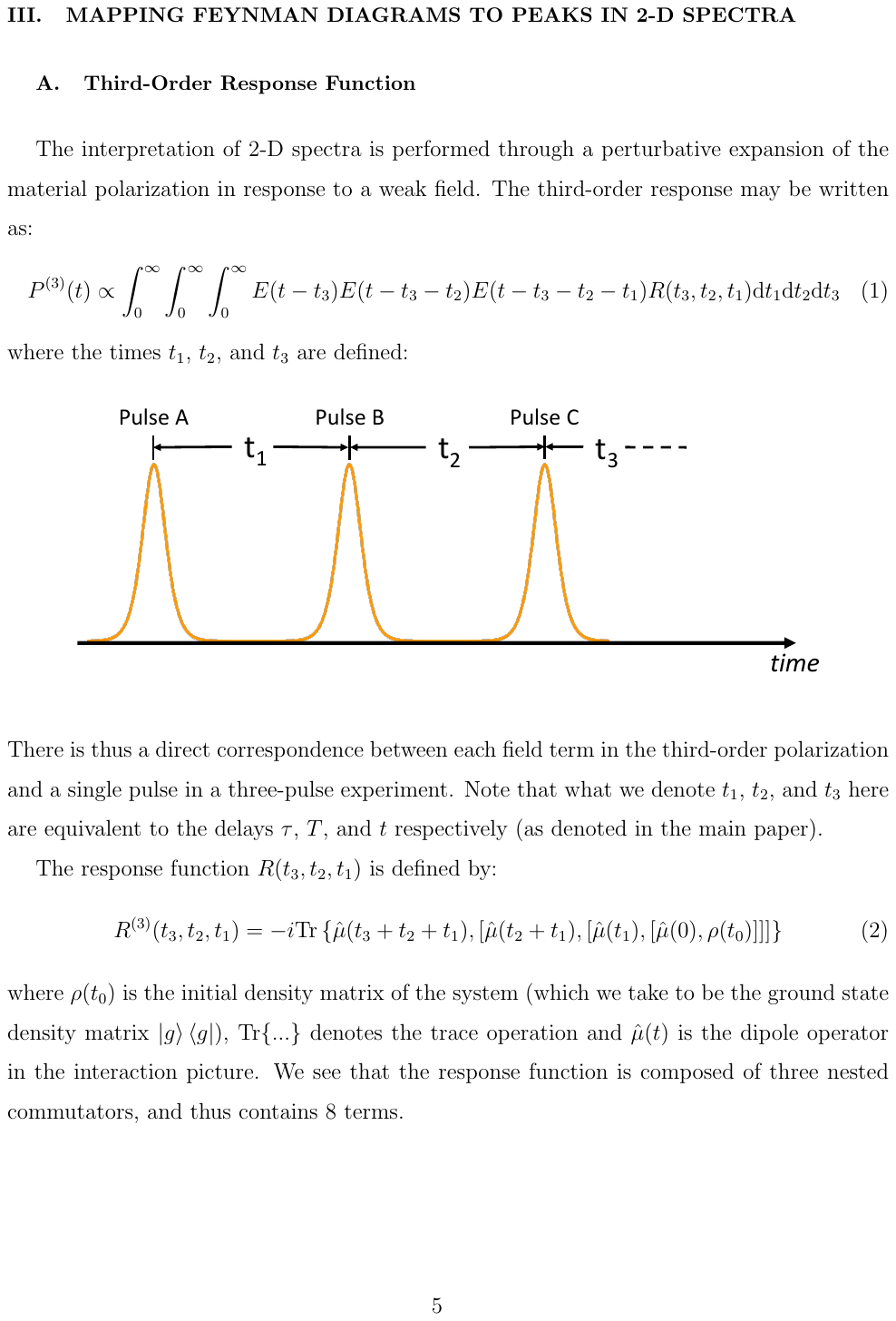}
\end{figure}
\begin{figure}
   \includegraphics[width=1\textwidth,page=6]{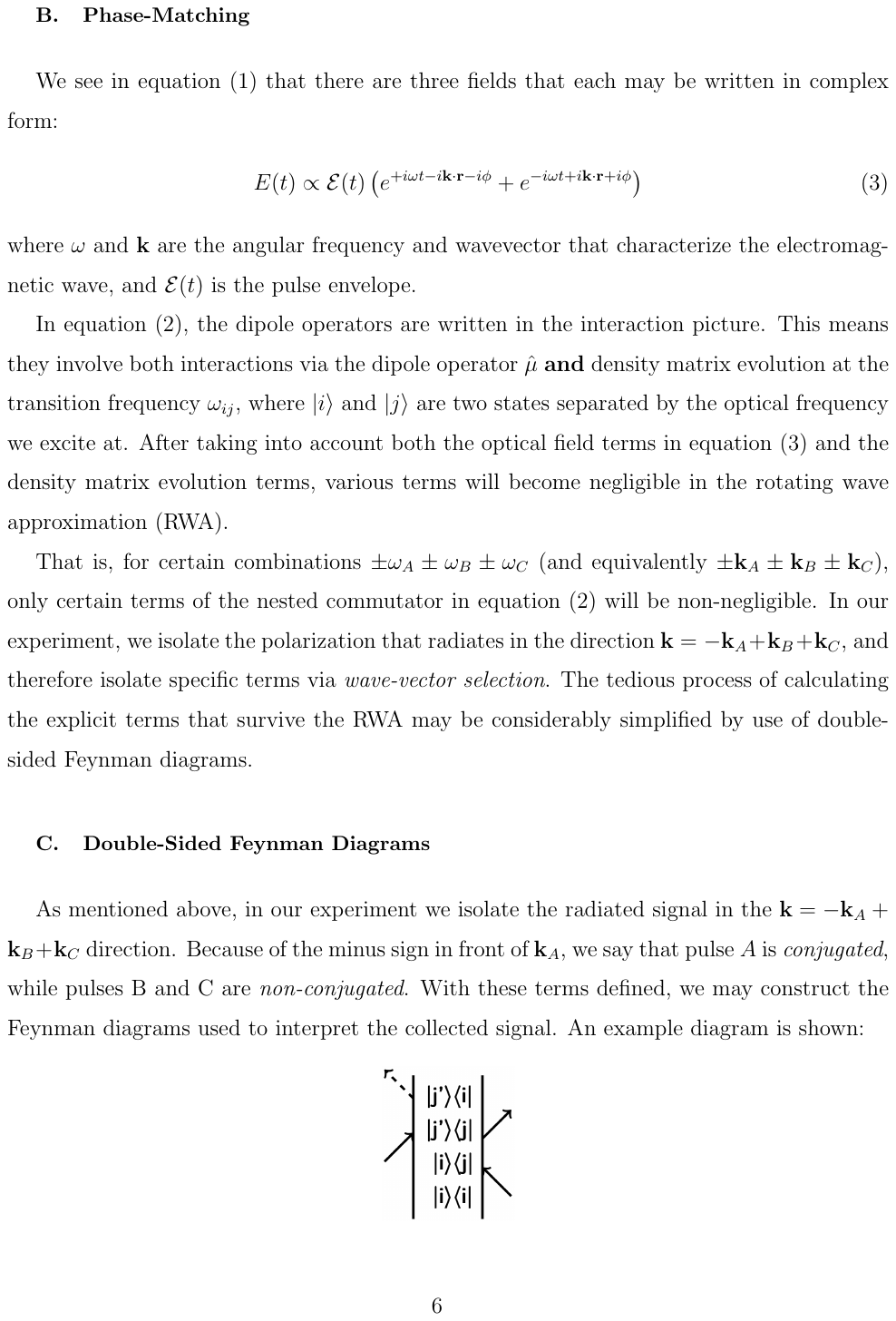}
\end{figure}
\begin{figure}
   \includegraphics[width=1\textwidth,page=7]{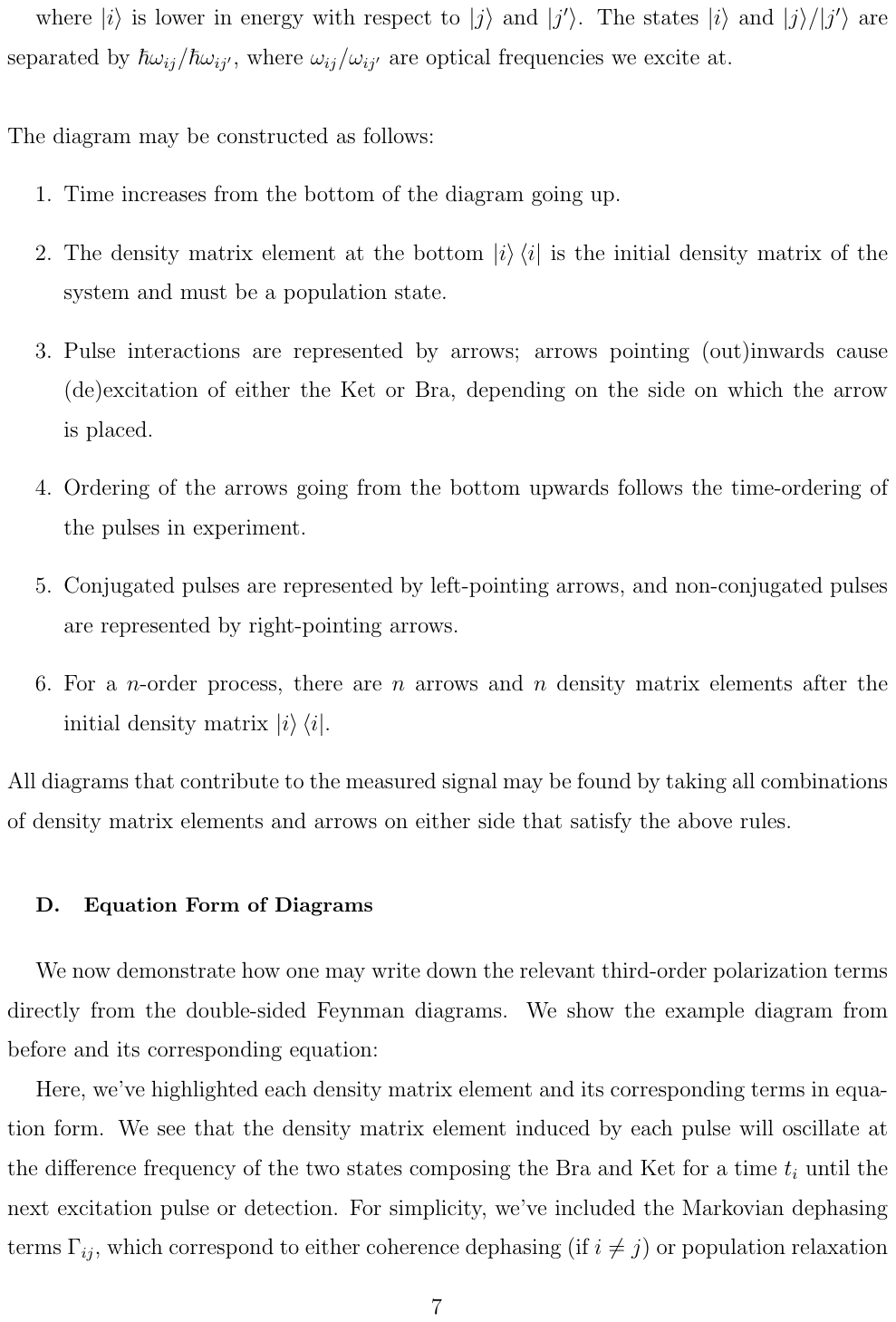}
\end{figure}
\begin{figure}
   \includegraphics[width=1\textwidth,page=8]{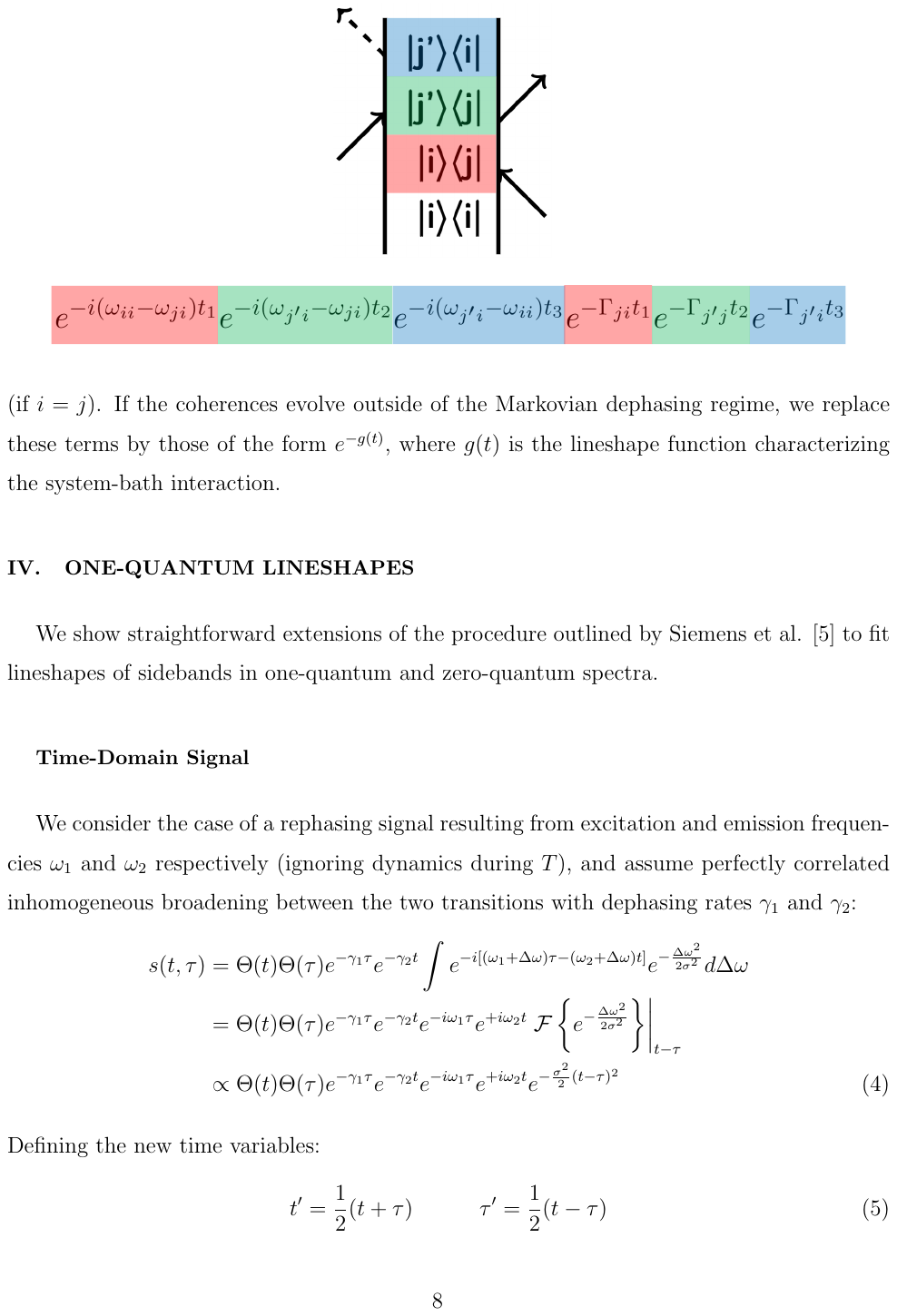}
\end{figure}
\begin{figure}
   \includegraphics[width=1\textwidth,page=9]{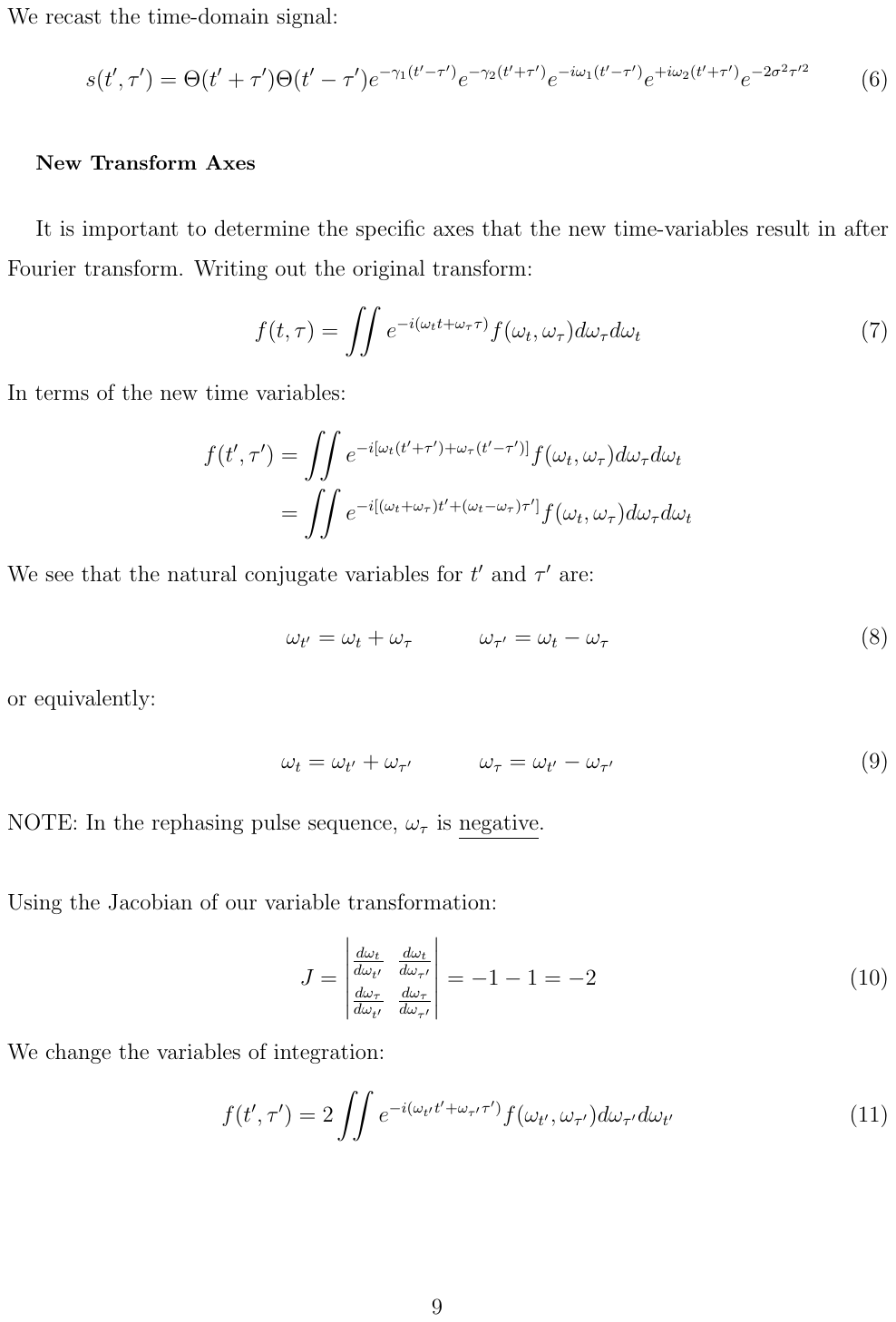}
\end{figure}
\begin{figure}
   \includegraphics[width=1\textwidth,page=10]{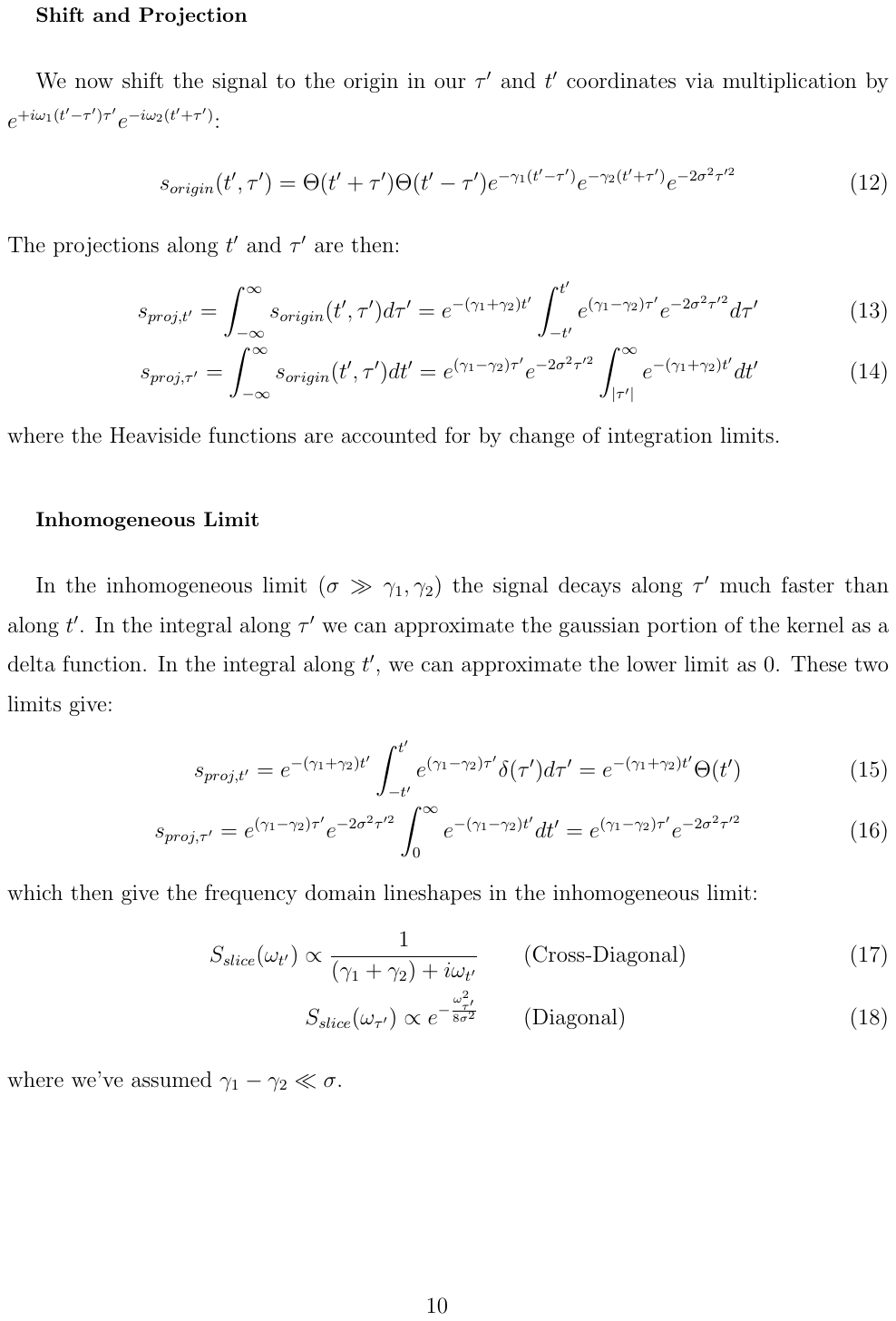}
\end{figure}
\begin{figure}
   \includegraphics[width=1\textwidth,page=11]{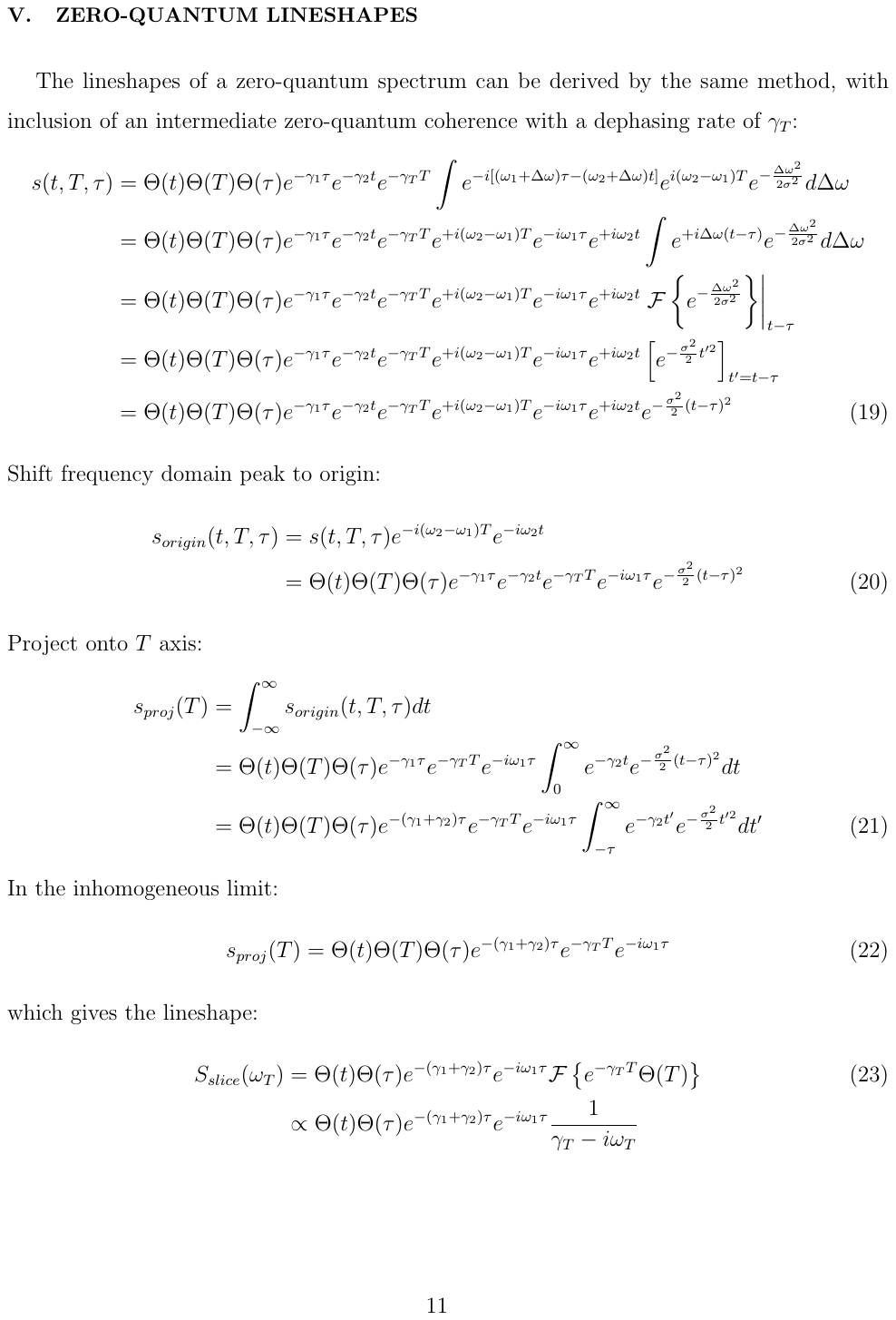}
\end{figure}
\begin{figure}
   \includegraphics[width=1\textwidth,page=12]{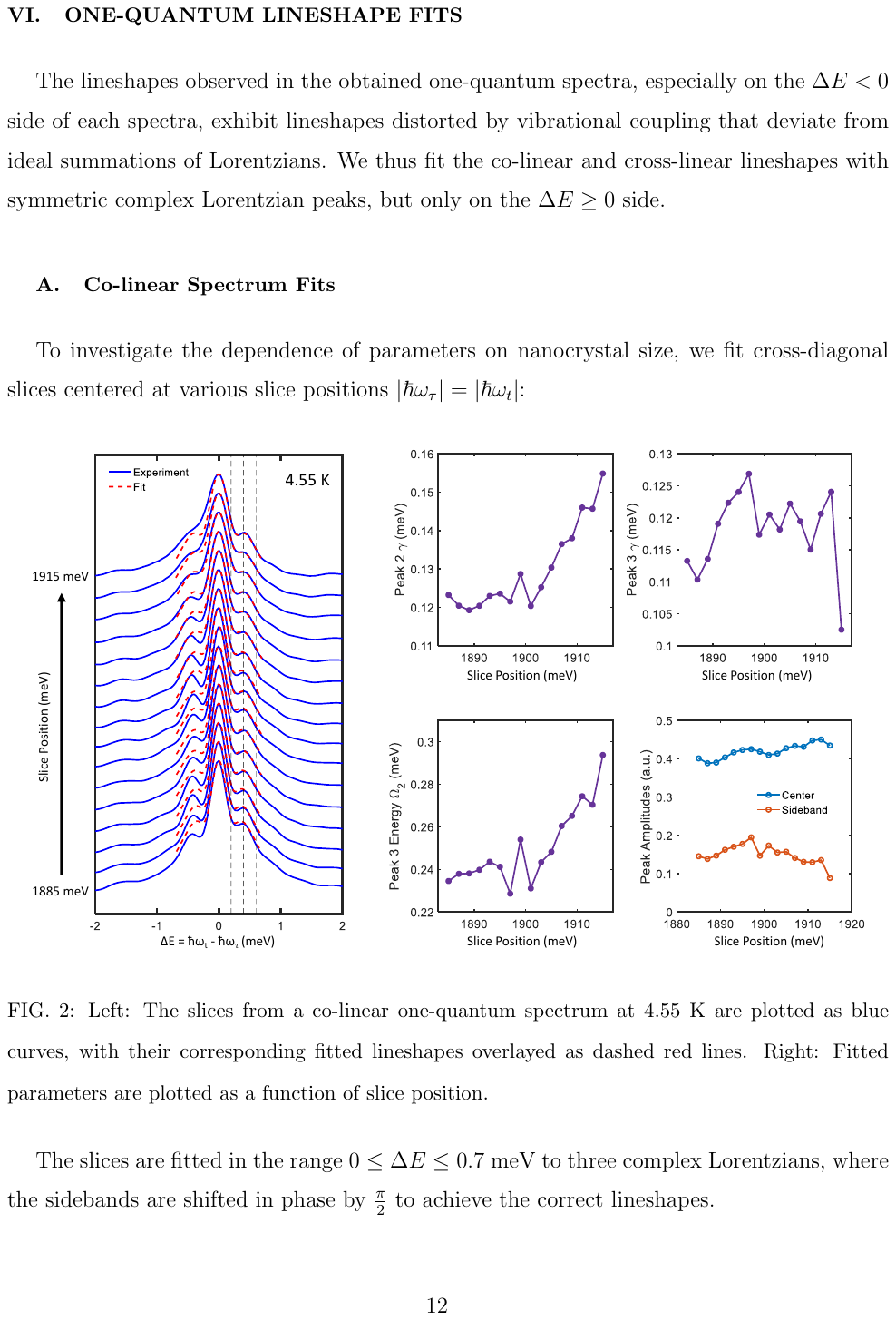}
\end{figure}
\begin{figure}
   \includegraphics[width=1\textwidth,page=13]{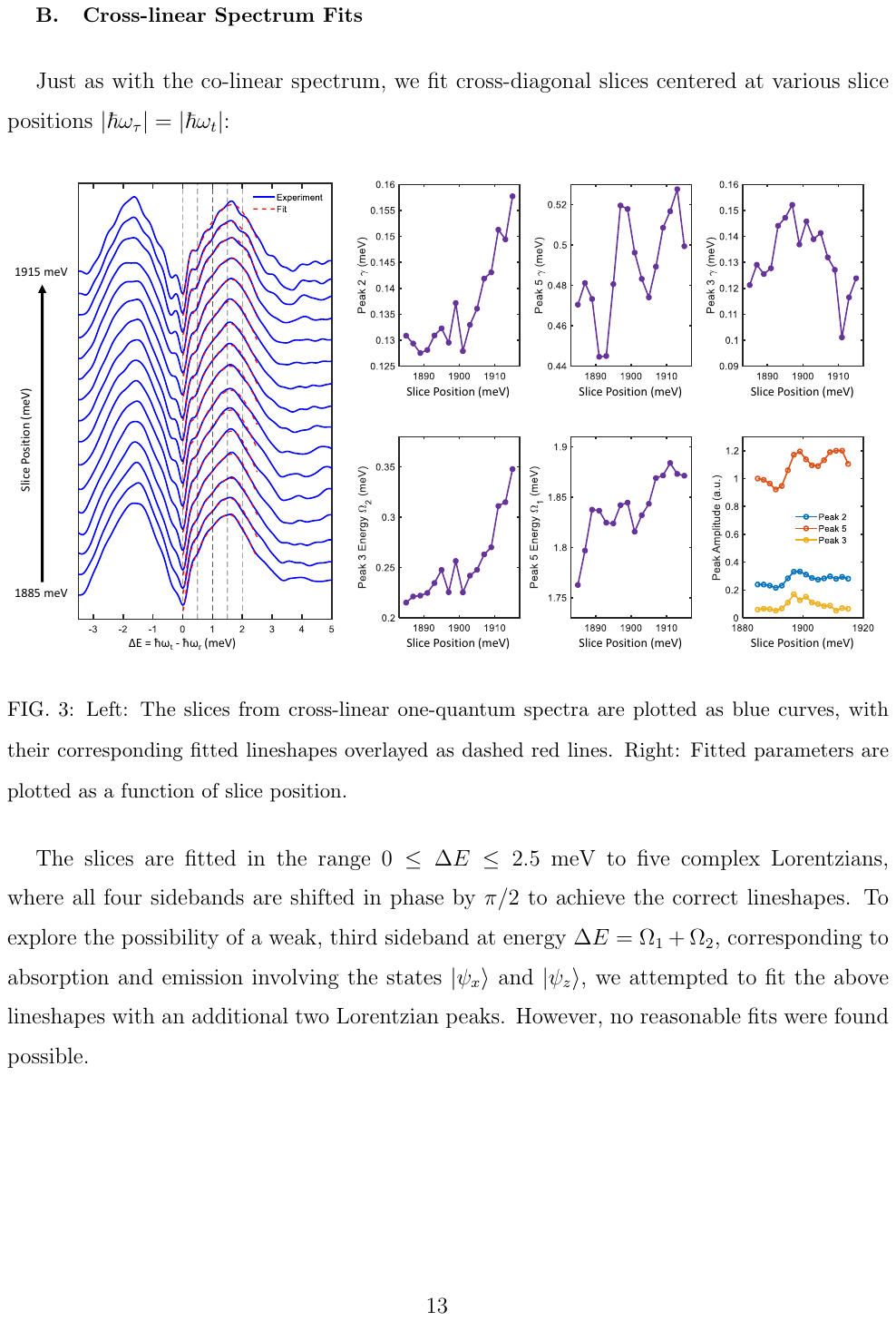}
\end{figure}
\begin{figure}
   \includegraphics[width=1\textwidth,page=14]{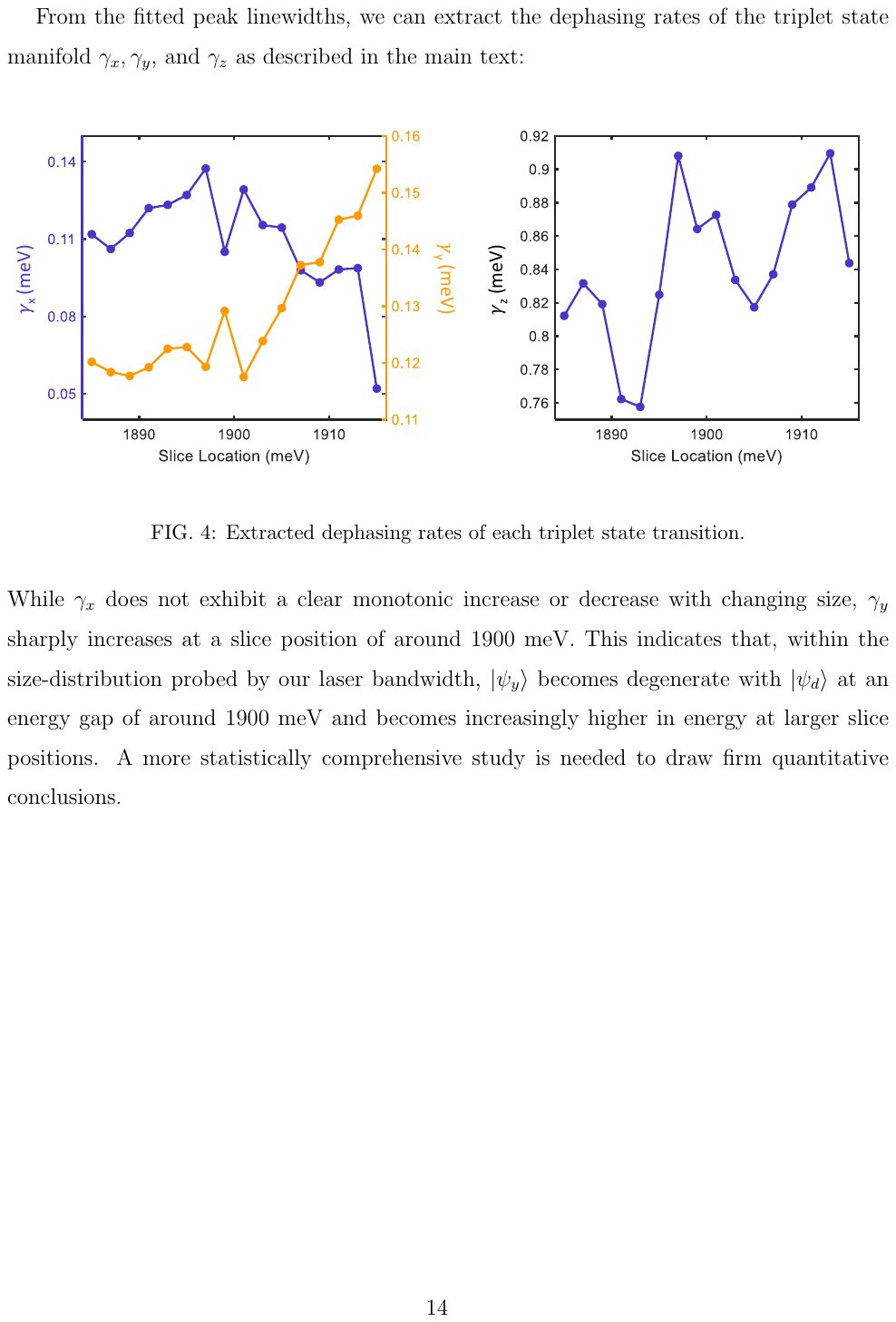}
\end{figure}
\begin{figure}
   \includegraphics[width=1\textwidth,page=15]{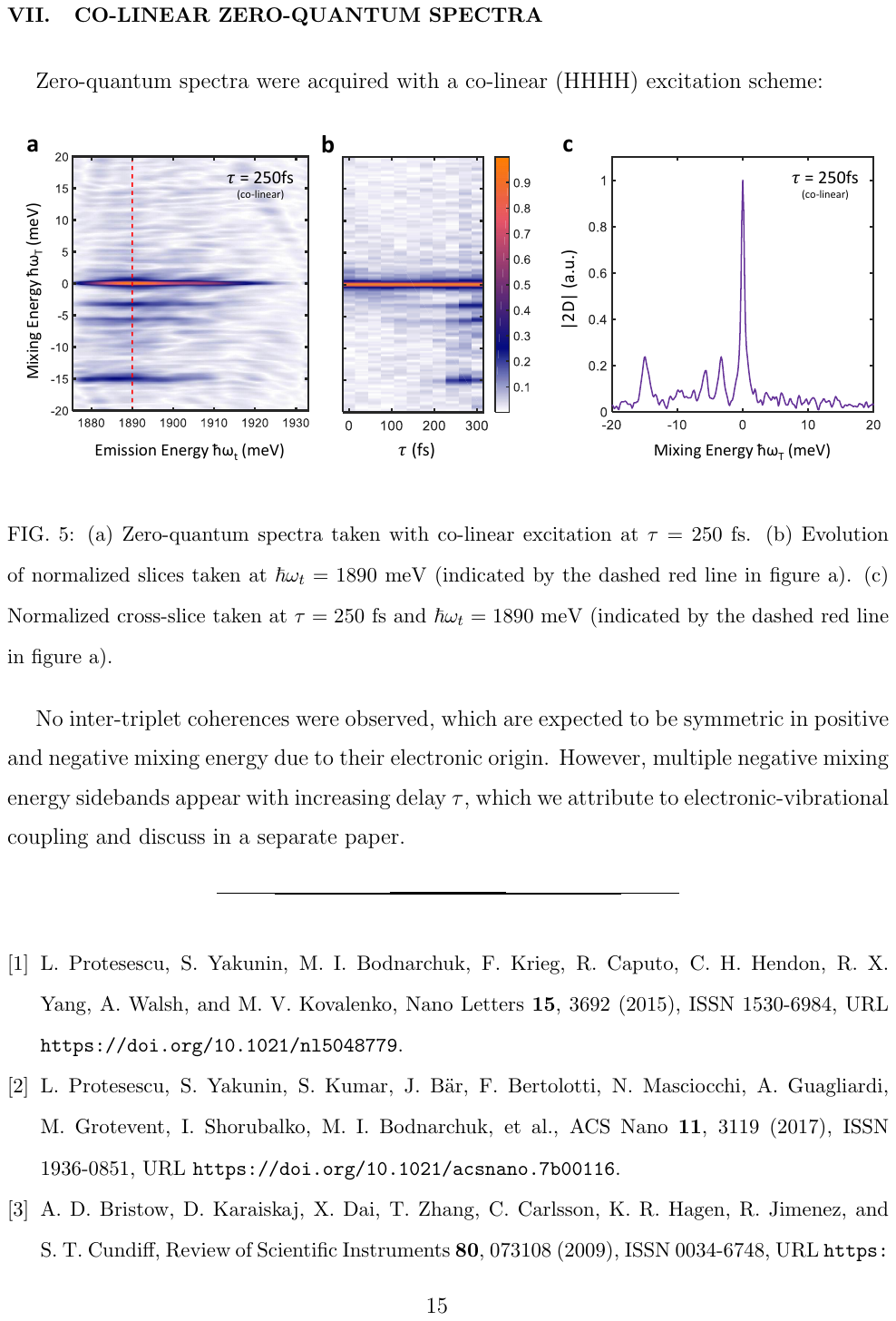}
\end{figure}
\begin{figure}
   \includegraphics[width=1\textwidth,page=16]{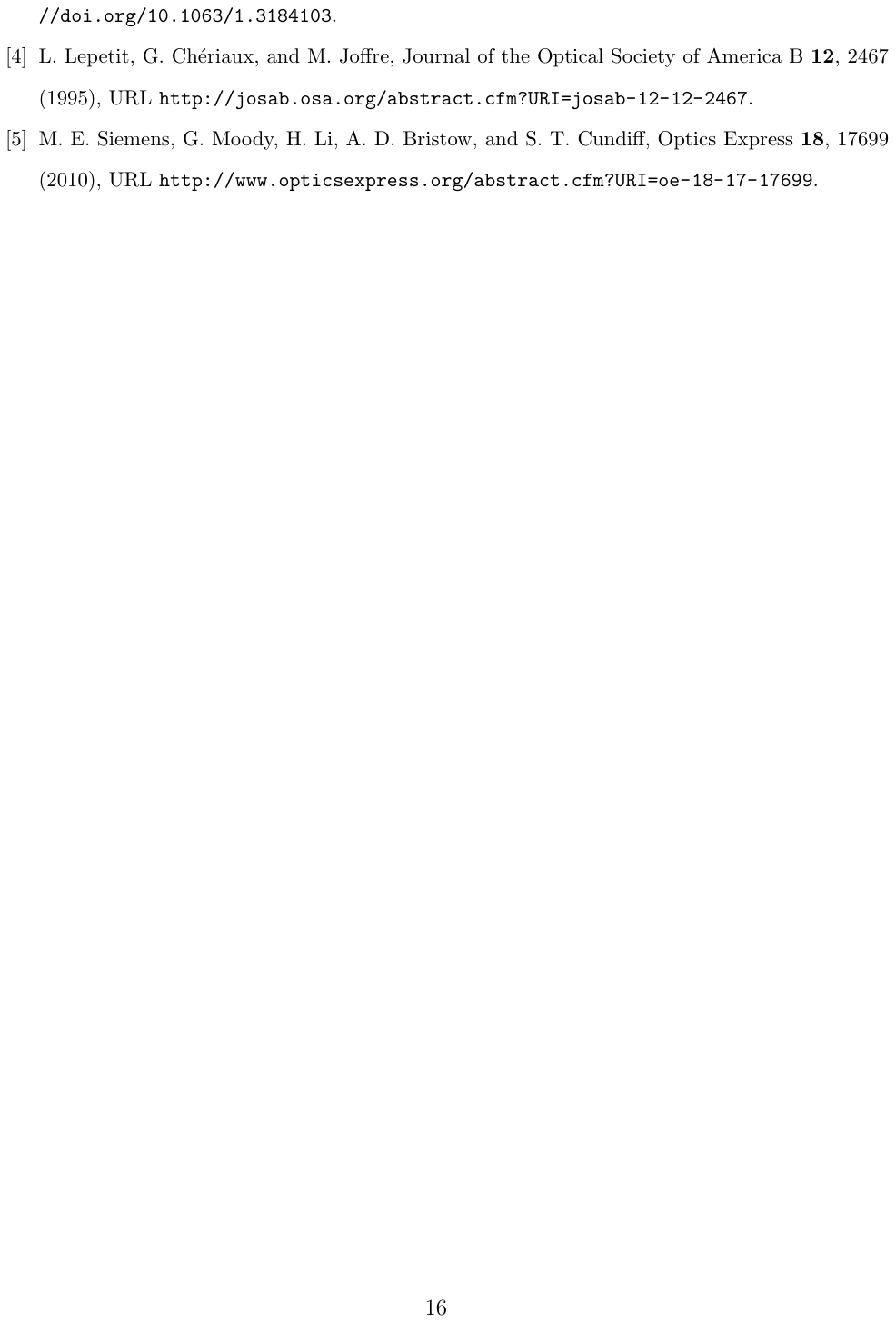}
\end{figure}

\end{document}